\documentclass[twoside]{dis08}
\usepackage[latin1]{inputenc}
\usepackage[dvips]{graphicx,epsfig,color}
\usepackage{wrapfig,rotating}
\usepackage{amssymb,amsmath,array}

\begin{document}
\title{Exclusive Diffractive Processes\\ within the Dipole Picture}

\author{G. Watt
%
\vspace{.3cm}\\
Department of Physics \& Astronomy, University College London, WC1E 6BT, UK
}

\maketitle

\begin{abstract}
  We discuss two different models for the impact parameter dependent dipole cross section: one based on DGLAP evolution and the other inspired by the Balitsky--Kovchegov equation.  The parameters are determined from fits to data on the total $\gamma^* p$ cross section measured at HERA.  The impact parameter dependent saturation scale is extracted.  Predictions are then confronted with HERA data on exclusive diffractive vector meson production and deeply virtual Compton scattering.  Finally, predictions are given for the cross sections of exclusive photoproduced $J/\psi$ and $\Upsilon$ mesons, and $Z^0$ bosons, expected at the Tevatron and LHC.
\end{abstract}

\section{Introduction}

The colour dipole model has proven to be very successful in describing a wide variety of small-$x$ inclusive and diffractive processes at HERA.  This talk~\cite{url} is based on work done on this subject in collaboration with H.~Kowalski and L.~Motyka \cite{Kowalski:2006hc,Watt:2007nr,Motyka:2008ac}.

\begin{wrapfigure}{r}{0.5\columnwidth}
    \centering
    \includegraphics[width=0.5\columnwidth]{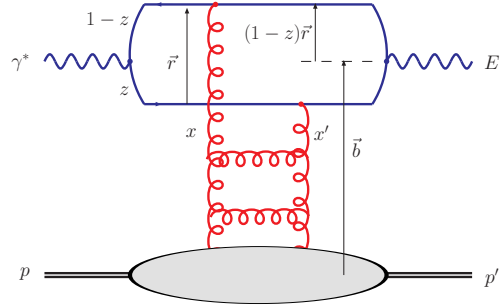}
  \caption{Diagram of the $\gamma^* p$ scattering amplitude.  For exclusive diffractive processes, such as vector meson production ($E=V$) or DVCS ($E=\gamma$), we have $x^\prime\ll x\ll 1$ and $t=(p-p^\prime)^2$.  For inclusive DIS, we have $E=\gamma^*$, $x^\prime=x\ll 1$ and $p^\prime=p$.}
  \vspace{-1cm}
  \label{fig:diagram}
\end{wrapfigure}
The amplitude for an exclusive diffractive process, $\gamma^* p\to E+p$, shown in Fig.~\ref{fig:diagram}, such as vector meson production, $E=V$, or deeply virtual Compton scattering (DVCS), $E=\gamma$, can be expressed as
\begin{multline} \label{eq:exclamp}
  \mathcal{A}^{\gamma^* p\rightarrow E+p}_{T,L}(x,Q,\Delta) = \mathrm{i}\,\int\!\mathrm{d}^2\boldsymbol{r}\int_0^1\!\frac{\mathrm{d}{z}}{4\pi}\int\!\mathrm{d}^2\boldsymbol{b}\;\\\times(\Psi_{E}^{*}\Psi)_{T,L}\;\mathrm{e}^{-\mathrm{i}[\boldsymbol{b}-(1-z)\boldsymbol{r}]\cdot\boldsymbol{\Delta}}\;\frac{\mathrm{d}\sigma_{q\bar q}}{\mathrm{d}^2\boldsymbol{b}},
\end{multline}
up to corrections from the real part of the amplitude and from skewedness ($x\ne x^\prime$).  Here, $z$ is the fraction of the photon's light-cone momentum carried by the quark, $r=|\boldsymbol{r}|$ is the transverse size of the $q\bar{q}$ dipole, while $\boldsymbol{b}$ is the impact parameter, that is, $b$ is the transverse distance from the centre of the proton to the centre-of-mass of the $q\bar{q}$ dipole; see Fig.~\ref{fig:diagram}.  The transverse momentum lost by the outgoing proton, $\boldsymbol{\Delta}$, is the Fourier conjugate variable to the impact parameter $\boldsymbol{b}$, and $t=-\Delta^2$.  The forward overlap function between the initial-state photon wave function and the final-state vector meson or photon wave function in \eqref{eq:exclamp} is denoted $(\Psi_E^*\Psi)_{T,L}$, while the factor $\exp[\mathrm{i}(1-z)\boldsymbol{r}\cdot\boldsymbol{\Delta}]$ in \eqref{eq:exclamp} originates from the non-forward wave functions.

Taking the imaginary part of the forward scattering amplitude immediately gives the formula for the total $\gamma^*p$ cross section:
\begin{equation}
  \sigma^{\gamma^* p}_{T,L}(x,Q) = {\rm Im}\,\mathcal{A}^{\gamma^* p\rightarrow \gamma^*p}_{T,L}(x,Q,\Delta=0) = \sum_f \int\!\mathrm{d}^2\boldsymbol{r} \int_0^1\!\frac{\mathrm{d} z}{4\pi}(\Psi^{*}\Psi)_{T,L}^f\,\int\!\mathrm{d}^2\boldsymbol{b}\;\frac{\mathrm{d}\sigma_{q\bar q}}{\mathrm{d}^2\boldsymbol{b}}. \label{eq:inclamp}
\end{equation}
The dipole picture therefore provides a unified description of both exclusive diffractive processes and inclusive DIS at small $x$.

\section{Impact parameter dependent dipole cross sections}
The unknown quantity common to \eqref{eq:exclamp} and \eqref{eq:inclamp} is the $b$-dependent dipole--proton cross section, 
\[
\frac{\mathrm{d}\sigma_{q\bar q}}{\mathrm{d}^2\boldsymbol{b}} = 2\;\mathcal{N}(x,r,b),
\]
where $\mathcal{N}$ is the imaginary part of the dipole--proton scattering amplitude, which can vary between zero and one, where $\mathcal{N}=1$ is the unitarity (``black disc'') limit.  The scattering amplitude $\mathcal{N}$ is generally parameterised according to some theoretically motivated functional form, with the parameters fitted to data.  We will consider two such parameterisations.

The first parameterisation is based on LO DGLAP evolution of an initial gluon density, $xg(x,\mu_0^2) = A_g\,x^{-\lambda_g}\,(1-x)^{5.6}$, with a Gaussian $b$ dependence, $T(b) \propto \mathrm{exp}(-b^2/2B_G)$.  We refer to this parameterisation as the ``b-Sat'' model~\cite{Kowalski:2006hc}:
\[
  \mathcal{N}(x,r,b) = 1-\exp\left(-\frac{\pi^2}{2N_c}r^2\alpha_S(\mu^2)\,xg(x,\mu^2)\,T(b)\right),
\]
where the scale $\mu^2=4/r^2+\mu_0^2$, $B_G=4$ GeV$^{-2}$ is fixed from the $t$-slope of exclusive $J/\psi$ photoproduction, and the other three parameters ($\mu_0^2$, $A_g$, $\lambda_g$) are fitted to ZEUS $F_2$ data with $x_{\mathrm{Bj}}\le 0.01$ and $Q^2\in[0.25,650]$ GeV$^2$.

The second parameterisation is a modified version of the colour glass condensate (CGC) dipole model of Iancu, Itakura and Munier:
\begin{equation} \label{eq:bcgc}
\mathcal{N}(x,r,b) =
\begin{cases}
  \mathcal{N}_0\left(\frac{rQ_s}{2}\right)^{2\left(\gamma_s+\frac{\ln(2/rQ_s)}{9.9\lambda \ln(1/x)}\right)} & :\quad rQ_s\le 2\\
  1-\mathrm{e}^{-A\ln^2(BrQ_s)} & :\quad rQ_s>2
\end{cases},
\end{equation}
where a Gaussian impact parameter dependence is introduced into the saturation scale:
\begin{equation} \label{eq:satbcgc}
  Q_s\equiv Q_s(x,b)=\left(\frac{x_0}{x}\right)^{\frac{\lambda}{2}}\;\left[\exp\left(-\frac{b^2}{2B_{\rm CGC}}\right)\right]^{\frac{1}{2\gamma_s}}.
\end{equation}
We refer to the parameterisation given by \eqref{eq:bcgc} and \eqref{eq:satbcgc} as the ``b-CGC'' model~\cite{Kowalski:2006hc,Watt:2007nr}.  The parameter $B_{\rm CGC}=7.5$ GeV$^{-2}$ is fixed from the $t$-slope of exclusive $J/\psi$ photoproduction, while the other four parameters ($\gamma_s$, $\mathcal{N}_0$, $x_0$, $\lambda$) are fitted to ZEUS $F_2$ data with $x_{\mathrm{Bj}}\le 0.01$ and $Q^2\in[0.25,45]$ GeV$^2$.  The optimum fitted value of the anomalous dimension at the saturation scale, $\gamma_s=0.46$, is close to the value of $\gamma_s\simeq 0.44$ determined from numerical solution of the Balitsky--Kovchegov equation.  However, the value of $\lambda = 0.119$ obtained from the fit is lower than the perturbatively calculated value of $\lambda\sim 0.3$, and suggests that the saturation scale comprises significant non-perturbative dynamics.

\begin{wrapfigure}{r}{0.4\columnwidth}
  \centering
  \includegraphics[width=0.4\columnwidth]{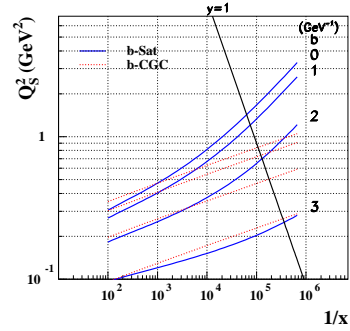}
  \caption{The impact parameter dependent saturation scale $Q_S^2\equiv 2/r_S^2$.}
    \label{fig:q2sx}
\end{wrapfigure}
It is customary to define a saturation scale $Q_S^2$, that is, the momentum scale at which the dipole--proton scattering amplitude $\mathcal{N}$ becomes sizable such that non-linear effects start to become important.  There is no unique definition of $Q_S^2$ and various choices are used in the literature.  We define the saturation scale $Q_S^2\equiv 2/r_S^2$, where the saturation radius $r_S$ is the dipole size where the scattering amplitude
\begin{equation} \label{eq:satdef}
  \mathcal{N}(x,r_S,b) = 1 - \mathrm{e}^{-\frac{1}{2}} \simeq 0.4.
\end{equation}
Note that we use lower-case $s$ and upper-case $S$ to distinguish between the two scales defined by \eqref{eq:satbcgc} and \eqref{eq:satdef} respectively.  The model-independent saturation scale $Q_S^2$ is shown in Fig.~\ref{fig:q2sx}: it is generally less than $0.5$ GeV$^2$ in the HERA kinematic regime for the most relevant impact parameters $b\sim 2$--$3$ GeV$^{-1}$.

\section{Exclusive diffractive processes at HERA}
\begin{wrapfigure}{l}{0.4\columnwidth}
  \centering
  \includegraphics[width=0.4\columnwidth]{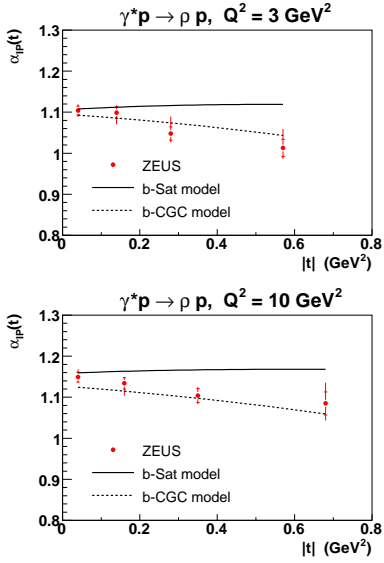}
  \caption{The effective Pomeron trajectory $\alpha_{\mathbb{P}}(t)$ vs.~$|t|$, where $\alpha_{\mathbb{P}}(t)$ is determined by fitting $\mathrm{d}\sigma/\mathrm{d} t\propto W^{4[\alpha_{\mathbb{P}}(t)-1]}$.}\vspace{-1cm}
    \label{fig:apom}
\end{wrapfigure}
\begin{figure}[ht]
  \centering
  \includegraphics[width=0.33\textwidth]{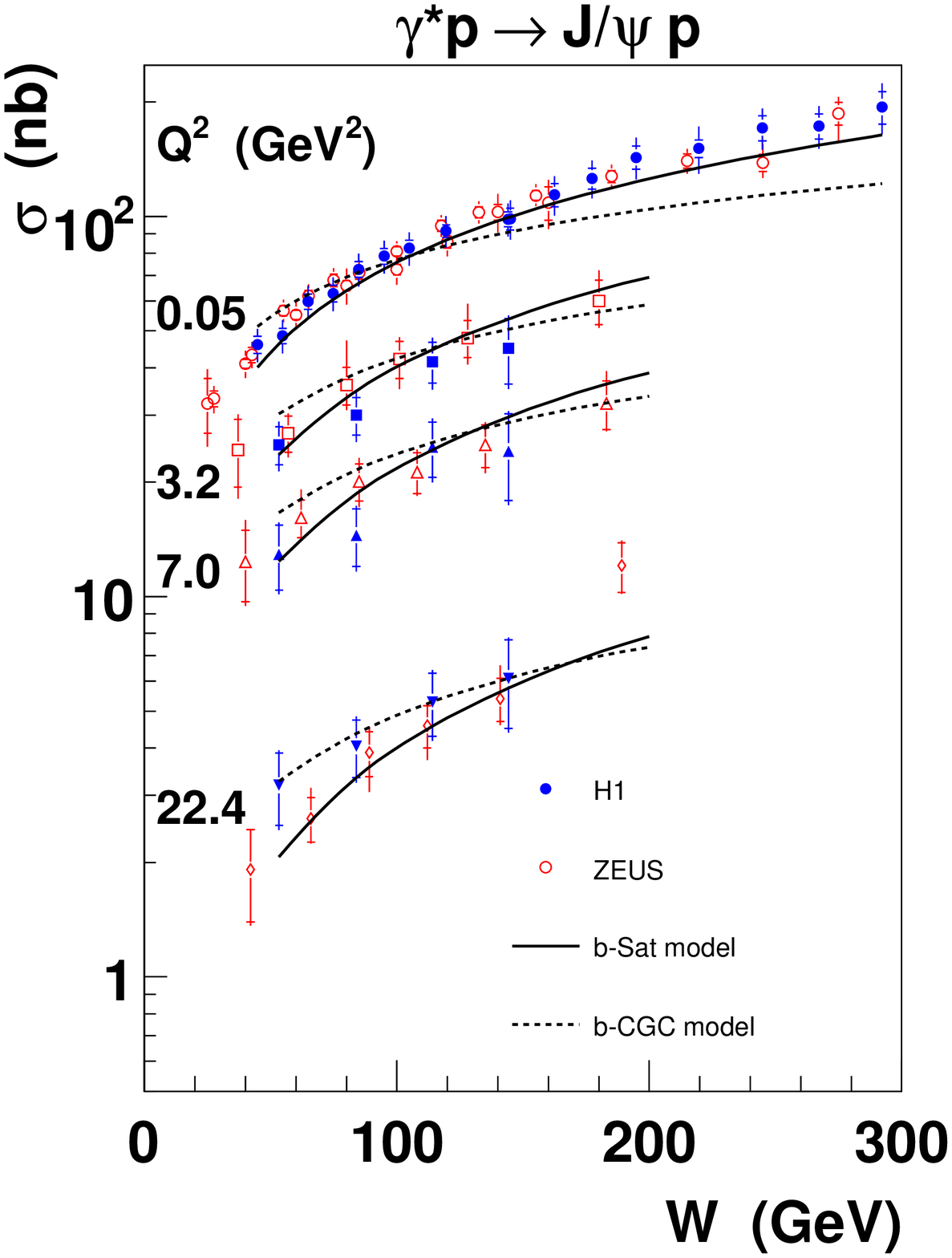}%
  \includegraphics[width=0.33\textwidth]{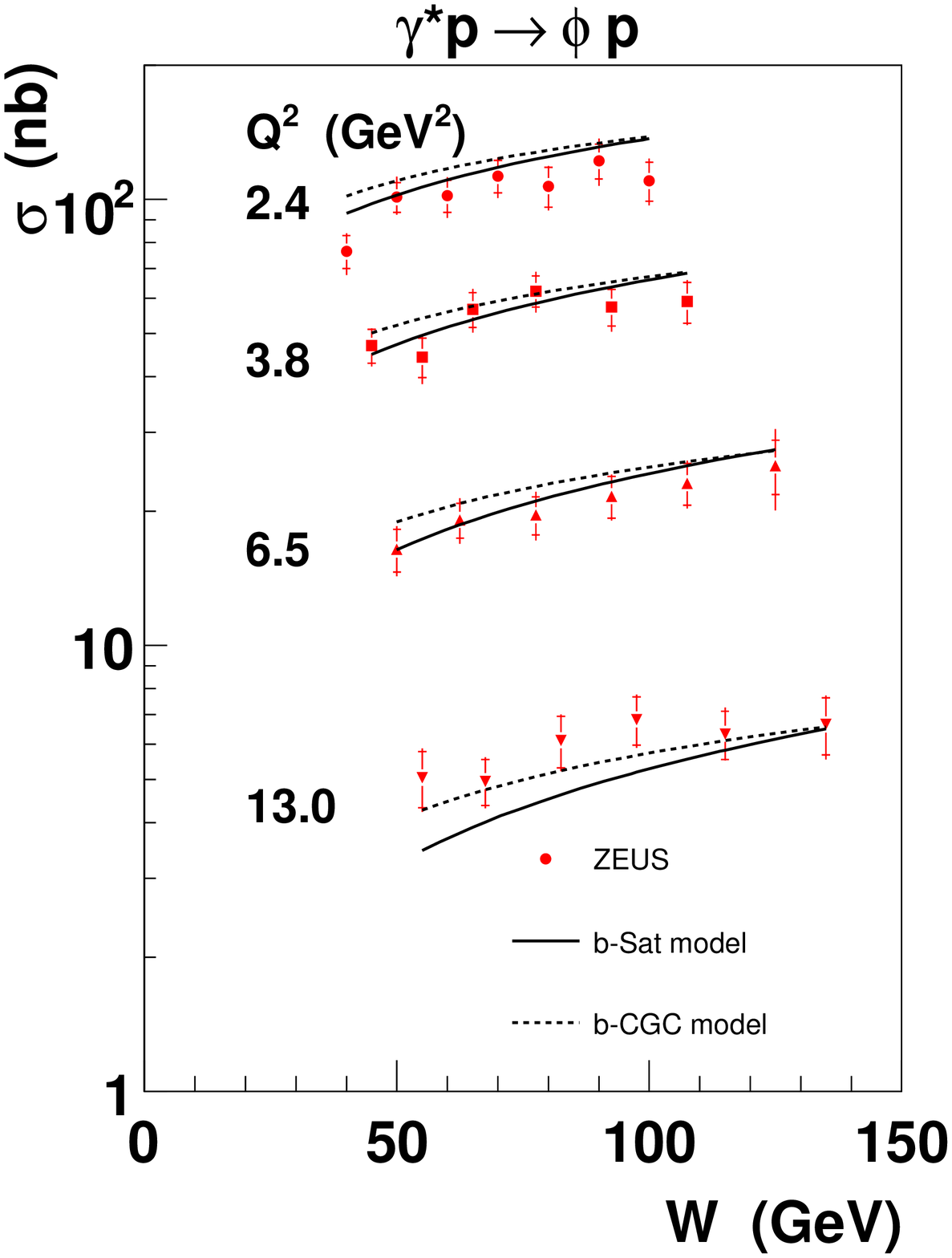}%
  \includegraphics[width=0.33\textwidth]{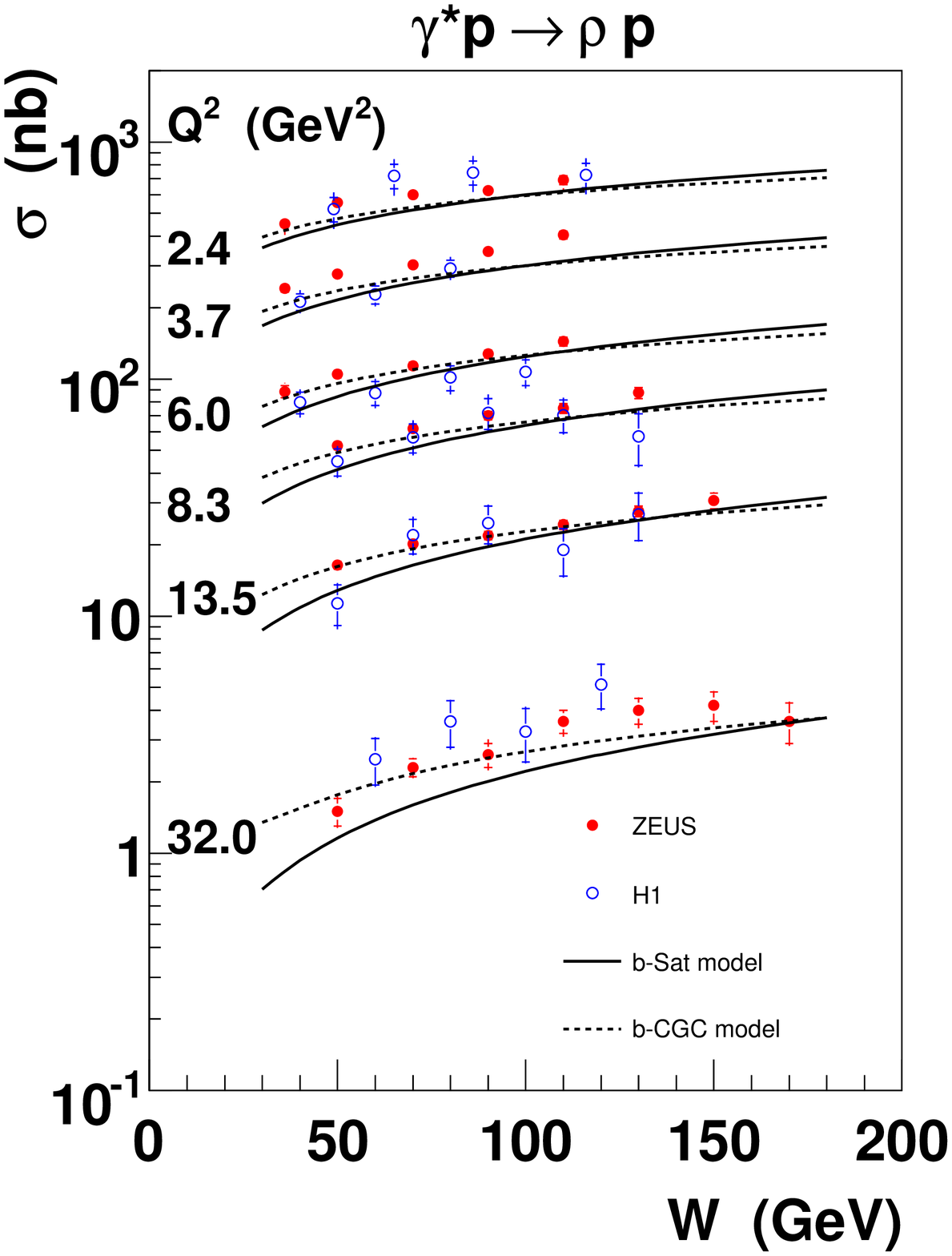}
  \caption{The total cross section $\sigma$ vs.~$W$ for exclusive $J/\psi$, $\phi$ and $\rho$ meson production compared to predictions from the b-Sat and b-CGC models~\cite{Watt:2007nr}.}\vspace{-0.2cm}
  \label{fig:crossw}
\end{figure}

A wealth of HERA data exists on exclusive diffractive vector meson ($\Upsilon$, $J/\psi$, $\phi$, $\rho$) production and DVCS.  It is therefore a significant challenge for an essentially parameter-free model to describe all aspects of the $Q^2$, $W$ and $t$ dependence.  Extensive comparison of the b-Sat and b-CGC model predictions with HERA data on exclusive processes has been made in Refs.~\cite{Kowalski:2006hc,Watt:2007nr}.  In general, both models describe almost all features of the available data.  Here we focus on only two aspects of the data which differentiate the models.  In Fig.~\ref{fig:apom} we show the effective Pomeron trajectory $\alpha_{\mathbb{P}}(t)$ vs.~$|t|$, where $\alpha_{\mathbb{P}}(t)$ is determined by fitting $\mathrm{d}\sigma/\mathrm{d} t\propto W^{4[\alpha_{\mathbb{P}}(t)-1]}$.  The b-CGC model gives a better description of $\alpha_{\mathbb{P}}^\prime$, where $\alpha_{\mathbb{P}}(t) = \alpha_{\mathbb{P}}(0)+\alpha_{\mathbb{P}}^\prime\,t$.  This suggests that the b-CGC dipole cross section better models the interplay between the $x$ and $b$ dependence.  In Fig.~\ref{fig:crossw} we show the $W$ dependence of the total cross sections for exclusive $J/\psi$, $\phi$ and $\rho$ meson production.  The $W$ dependence of $J/\psi$ photoproduction is much better described by the b-Sat model than by the b-CGC model.  This difference can be traced to relatively small differences in the form of the dipole cross sections.

\newpage

\section{Exclusive photoproduction at the Tevatron and LHC}
\vspace*{-0.2cm}
\begin{wraptable}{r}{0.45\columnwidth}
  \centering
  \begin{tabular}{|c|c|c|}
    \hline
    $J/\psi$ & $\mathrm{d}\sigma/\mathrm{d} y|_{y=0}$ (nb) & $\sigma$ (nb) \\ \hline
    Tevatron & $3.4$ & $28$ \\
    LHC & $9.8$ & $120$ \\
    \hline\multicolumn{3}{c}{}\\\hline
    $\Upsilon(1S)$ & $\mathrm{d}\sigma/\mathrm{d} y|_{y=0}$ (pb) & $\sigma$ (pb) \\ \hline
    Tevatron & $14$ & $115$ \\
    LHC & $72$ & $1060$ \\
    \hline\multicolumn{3}{c}{}\\\hline
    $Z^0$ & $\mathrm{d}\sigma/\mathrm{d} y|_{y=0}$ (fb) & $\sigma$ (fb) \\ \hline
    Tevatron & $0.077$ & $0.30$ \\
    LHC & $1.4$ & $13$ \\
    \hline
  \end{tabular}
  \caption{``b-Sat'' model predictions~\cite{Motyka:2008ac} for $J/\psi$, $\Upsilon$ and $Z^0$ photoproduction.}\vspace{-0.2cm}
  \label{tab:results}
\end{wraptable}

The equivalent-photon approximation allows predictions to be made for the rapidity distributions of exclusive photoproduced $J/\psi$ and $\Upsilon$ mesons, and $Z^0$ bosons, expected at the Tevatron and LHC.  A given rapidity $y$ corresponds to a photon energy $k\simeq (M_E/2)\exp(y)\simeq W^2/(2\sqrt{s})$.  The hadron--hadron cross sections are obtained from the photon--hadron cross sections by multiplying by the flux $\mathrm{d} n/\mathrm{d} k$ of quasi-real photons.  We neglect the possible interference between photon--Pomeron and Pomeron--photon fusion.  We also neglect absorptive corrections from soft rescattering.  Both these effects should be largely washed out for cross sections integrated over final state momenta, in which case the rapidity gap survival factor is expected to be $S^2\sim 0.7$--$0.9$.  In Table \ref{tab:results} we present $\mathrm{d}\sigma/\mathrm{d} y$ at $y=0$ and the total cross sections integrated over rapidity.  The $J/\psi$ and $\Upsilon$ predictions have been scaled by factors 1.08 and 2.96, respectively, in order to give the best agreement with the existing HERA data.


\begin{footnotesize}

\end{footnotesize}


\end{document}